\documentclass[11pt]{article}
\usepackage{graphicx,color,bm}
\usepackage{fancyhdr,lastpage}
\usepackage{subeqn}
\usepackage{slashed}
\def\simgt{\,{\rlap{\lower 3.5pt\hbox{$\mathchar\sim$}}\raise 1pt\hbox{$>$}}\,}
\def\simlt{\,{\rlap{\lower 3.5pt\hbox{$\mathchar\sim$}}\raise 1pt\hbox{$<$}}\,}
\def\fds{f_{D_s}}

\def\dstaunu{D_s \to \tau \nu}
\def\bptaunu{B^+ \to \tau \nu}
\def\rds{r_{D_s}}
\def\rbp{r_{B^+}}

\begin{document}
\thispagestyle{empty}
\vspace*{-15mm}
\baselineskip 10pt
\begin{flushright}
\begin{tabular}{l}
{\bf OCHA-PP-307}\\
\today
\end{tabular}
\end{flushright}
\baselineskip 24pt 
\vglue 10mm 
\begin{center}
{\Large\bf
Constraints on $R$-parity violating interactions \\
in supersymmetric standard model \\
from leptonic decays of $D_s$ and $B^+$ mesons
}
\vspace{5mm}

\baselineskip 18pt
\def\thefootnote{\fnsymbol{footnote}}
\setcounter{footnote}{0}
{\bf
Gi-Chol Cho$^{1)}$ and Hikaru Matsuo$^{2)}$
\vspace{5mm}

$^{1)}${\it 
Department of Physics, Ochanomizu University, Tokyo 112-8610, Japan
}\\
$^{2)}${\it 
Graduate School of Humanities and Sciences,
Ochanomizu University, Tokyo, 112-8610, Japan
}\\
}
\vspace{10mm}

\end{center}

\begin{center}
{\bf Abstract}\\[7mm]
\begin{minipage}{14cm}
\baselineskip 18pt
\noindent
We study leptonic decays $D_{s}\to \tau \nu$ and $B^+ \to \tau \nu$ 
In $R$-parity violating (RPV) supersymmetric standard model. 
The interference between the $s$-channel slepton exchange and 
the  $t$-channel squark exchange diagrams could be destructive 
in a certain model parameter region so that sizable RPV couplings 
are allowed. 
Contributions from the RPV interactions to the final states with 
the lepton flavor violation (LFV), $\tau\nu_\mu (\tau \nu_e)$, are also 
examined taking account of constraints on the RPV couplings from the 
other LFV processes.  

\end{minipage}
\end{center}
\newpage

\baselineskip 18pt
\section{Introduction}
Although the standard model (SM) of particle physics has shown a good  
agreement with almost all the experimental results (for example, 
see~\cite{:2005ema}), the gauge hierarchy problem motivates us to 
explore physics beyond the SM. 
Supersymmetry (SUSY) has been expected as a promising idea to solve 
the gauge hierarchy problem~\cite{Martin:1997ns}, and looking for 
direct or indirect signatures of supersymmetric extension of the 
SM is one of the most important tasks of collider experiments at energy
frontier such as LHC. 
%

It is well known that, in the supersymmetric SM, the baryon ($B$) and
lepton ($L$) numbers are not conserved in general. 
The $R$-parity is, therefore, introduced to protect the proton longevity  
or to suppress unobserved $B$- or $L$-number violating processes.  
An important consequence of $R$-parity is that the lightest
supersymmetric particle (LSP) is stable and it could be a candidate
of the cold dark matter. 
On the other hand, if some of the $R$-parity violating (RPV) 
interactions are allowed, phenomenology of supersymmetric SM is 
drastically changed. 
For example, tiny neutrino mass can be explained through the RPV 
interactions without introducing a heavy Majorana 
neutrino~\cite{Hall:1983id}.  
Even without $R$-parity, the light gravitino could play a role of cold 
dark matter~\cite{Buchmuller:2007ui}.  
%

%
In ref.~\cite{Aida:2010xi}, we studied the leptonic decays  
$D_s \to \tau \nu$ and $B^+ \to \tau \nu$ 
in the RPV supersymmetric SM. 
It was pointed out that the experimental data of
$\dstaunu$~\cite{Naik:2009tk} shows 
the deviation from the Lattice QCD calculations about  
2.4$\sigma$~\cite{Follana:2007uv},  
while the discrepancy between the experimental measurement of 
$\bptaunu$ and 
the SM expectation is about 2.5$\sigma$~\cite{Bona:2009cj}. 
In the supersymmetric SM with RPV interactions, the leptonic decays
occur at tree level through  
(i) charged slepton exchange in the $s$-channel diagram, and  
(ii) down-squark exchange in the $t$-channel diagram. 
We found that the supersymmetric contributions could either 
constructively or destructively interfere  
between the $s$- and $t$-channel diagrams so that the $R$-parity 
couplings could be sizable when the contributions are cancelled 
each other which has not been pointed out in earlier 
works~\cite{Baek:1999ch,Dreiner:2001kc,Dreiner:2006gu,  
Kundu:2008ui, Kao:2009mz, Bhattacharyya:2009hb}. 
Although the RPV interactions lead to the final states where the 
charged lepton and neutrino are not only flavor diagonal 
($\tau\nu_\tau$) but also off-diagonal ($\tau \nu_\mu$ or $\tau \nu_e$),  
the latter was neglected in the previous works\footnote{
The flavor violating final states in the leptonic decay of $D_s$ meson 
have been discussed in the framework of leptoquark models 
in ref.~\cite{Dorsner:2009cu}. 
}. 
%

%
Recently, the Lattice QCD calculation of the decay constant $f_{D_s}$ 
has been updated~\cite{Davies:2010ip} and the world average of the 
experimental data of $\dstaunu$ is smaller than 
the previous one~\cite{hfag:fds}. 
As a result, the deviation between the SM expectation and the 
experimental data of $\dstaunu$ decreased from 2.4$\sigma$ to 
1.6$\sigma$ and so called the ``$\fds$-puzzle'' may disappear. 
%

%
In this paper, we study constraints on the RPV interactions from 
$\dstaunu$ and $\bptaunu$ taking account of the recently reported 
result on $f_{D_s}$.  
Although the deviation in $\dstaunu$ ($\fds$) is now 1.6$\sigma$, 
we show that there are still sizable allowed parameter space of the 
RPV couplings where the supersymmetric contributions are cancelled 
between the $s$- and $t$-channel diagrams. 
We also study in this paper the contributions of the RPV interactions 
to the leptonic decays with the lepton flavor violation (LFV) 
($D_s, B^+ \to \tau\nu_i$, $i=e,\mu$). 
The RPV interactions which lead to the LFV in the final state of the 
leptonic decays also contribute to the other LFV process, such as 
$\tau \to \mu\gamma$ and $\tau \to \mu\eta$ for $\dstaunu$, and 
$B^+ \to \pi^+ \nu \bar{\nu}$ and $B^0 \to \ell_i^\pm \ell_j^\mp$ for
$\bptaunu$. 
We show that contributions of RPV interactions to 
$D_s \to \tau \nu_\mu$ are suppressed by a few order of magnitude 
compared to the flavor diagonal case ($\tau\nu_\tau$) 
owing to the constraints from the other LFV processes, while those 
for $B^+$ can be as large as $\tau\nu_\tau$. 
%

%
This paper is organized as follows. In the next section, we briefly
review the relevant RPV interactions for the leptonic decays of $D_s$ 
and $B^+$ mesons. 
The numerical results are presented in Sec.~\ref{numerical}. 
Sec.~\ref{summary} is devoted to summary and discussions. 
\section{Set up}\label{setup}
The trilinear interactions with $R$-parity violations 
are described by the following superpotential 
\begin{eqnarray}
W_{\slashed R} &=& 
\frac{1}{2}\lambda_{ijk} L_i L_j E_k 
+ \lambda'_{ijk} L_i Q_j D_k 
+ \frac{1}{2} \lambda''_{ijk} U_i D_j D_k,  
\label{rparity}
\end{eqnarray}
where $Q$ and $L$ are SU(2)$_L$ doublet quark and lepton superfields, 
respectively. 
The up- and down-type singlet quark superfields are represented by 
$U$ and $D$, while the lepton singlet superfield is $E$. 
The generation indices are labeled by $i,j$ and $k$. 
The SU(2)$_L$ and SU(3)$_C$ gauge indices are suppressed. 
The dimensionless coefficient $\lambda_{ijk}$ is anti-symmetric for 
$i$ and $j$, while $\lambda''_{ijk}$ is anti-symmetric for $j$ and $k$.  
For a comprehensive review of the $R$-parity violating supersymmetric
SM, see, ref.~\cite{Barbier:2004ez}. 
Constraints on the RPV couplings 
$\lambda_{ijk}, \lambda'_{ijk}$ and $\lambda''_{ijk}$ from various 
processes have been studied in the literature~
\cite{Barger:1989rk, Dreiner:1997uz, Bhattacharyya:1997vv, 
Allanach:1999ic, Dreiner:2006gu}. 
Since the baryon number violating coupling $\lambda''_{ijk}$ 
induces too fast proton decay, we take $\lambda''_{ijk}=0$ in 
the following.  
Then, the leptonic decays of $D_s$ and $B^+$ mesons occur through 
the slepton exchange in the $s$-channel diagram with a product of
$\lambda$ and $\lambda'$, and the squark exchange in the $t$-channel 
diagram with a product of two $\lambda'$ couplings.  

%
Let us briefly summarize the leptonic decay of a pseudo scalar meson $P$ 
which consists of the up and (anti-) down-type quarks $u_a$ and  
$\bar{d_b}$, where $a,b$ are generation indices of quarks. 
The decay width of $P \to \ell_i \nu_j$ is given as 
\begin{eqnarray}
\Gamma(P \to \ell_i \nu_j ) 
&=& \frac{1}{8\pi} r_P^2 G_F^2 |V_{u_a d_b}^*|^2 f_P^2 m_{\ell_i}^2
m_P \left(1-\frac{m_{\ell_i}^2}{m_P^2}\right)^2
\label{dw}
\end{eqnarray}
where $G_F,V_{u_a d_b},m_{\ell_i}$ and $m_P$ are the Fermi constant, the 
Cabibbo-Kobayashi-Maskawa matrix element, the mass of a charged lepton 
$\ell_i$ and the mass of a pseudo scalar meson $P$, 
respectively. 
The flavor indices of charged leptons and neutrinos are expressed 
by $i$ and $j$, respectively. 
The decay constant is denoted by $f_P$. 
A parameter $r_P$ is defined as, 
\begin{eqnarray}
r_P^2 &\equiv& \frac{\left|G_F V_{u_a d_b}^* + A^P_{ii} \right|^2}
{ G_F^2 \left|V_{u_a d_b}^*\right|^2}
+
\sum_{j(\neq i)}
\frac{\left|A^P_{ij} \right|^2}
{ G_F^2 \left|V_{u_a d_b}^*\right|^2}, 
\label{rparam}
\end{eqnarray}
where $A^P_{ij}$ represents new physics contribution. 
Note that, in the second term of r.h.s. in eq.~(\ref{rparam}), 
one should take a sum only for $j$ (neutrinos), because that 
the neutrino flavor cannot be detected experimentally. 
If there is no new physics contribution, $r_P=1$. 
The interaction Lagrangian of the $t$-channel contribution 
to the decay width (\ref{dw}) can be obtained from 
the superpotential (\ref{rparity});  
\begin{eqnarray}
{\cal L} &=& 
\lambda'_{ijk} \left\{
-\overline{(\ell^c_L)_i} (u_L)_j (\widetilde{d}_R)^*_k
\right\}
+ 
\lambda'_{ijk} \left\{
\overline{(\nu^c_L)_i} (d_L)_j
(\widetilde{d}_R)_k^*
\right\} + {\rm h.c.}. 
\label{tch}
\end{eqnarray} 
Using the Fierz transformation, 
the effective Lagrangian which describes the $t$-channel  
squark exchange is given as 
\begin{eqnarray}
{\cal L_{\rm eff}^{\it t}} &=& 
\frac{1}{8}\sum_{k=1}^3
\frac{\lambda'_{iak} \lambda'^*_{jbk}}
{m_{\widetilde{d}_{Rk}}^2} ~
\bar{\nu}_j \gamma^\mu (1-\gamma_5) \ell_i~
\bar{d}_b \gamma_\mu (1-\gamma_5) u_a. 
\end{eqnarray}
For comparison, we show the effective Lagrangian for the $W$-boson 
exchange 
\begin{eqnarray}
{\cal L_{\rm eff}^{\rm SM}} &=& 
\frac{G_F}{\sqrt{2}} V_{u_a d_b}^* 
\bar{\nu}_i \gamma^\mu (1-\gamma_5) \ell_i~
\bar{d}_b \gamma_\mu (1-\gamma_5) u_a. 
\label{wboson}
\end{eqnarray}
Using the decay constant $f_P$ which is defined by 
\begin{eqnarray}
\langle 0 | \bar{d}_b \gamma^\mu \gamma_5 u_a| P(q) \rangle 
= i f_P q^\mu, 
\label{decayconst}
\end{eqnarray}
we find the $t$-channel squark contribution to 
the decay $P(u_a \bar{d}_b) \to \ell_i \nu_j$ as  
\begin{eqnarray}
(A_t^P)_{ij} &=& \frac{1}{4\sqrt{2}}
\sum_{k=1}^3
\frac{\lambda'_{iak} \lambda'^*_{jbk}}
{m_{\widetilde{d}_{Rk}}^2}. 
\label{atp}
\end{eqnarray}
The $s$-channel contribution can be calculated from the interaction 
Lagrangian 
\begin{eqnarray}
{\cal L} &=& \lambda_{ijk} 
\left\{-\overline{(\ell_R)_k} (\nu_L)_j (\widetilde{\ell}_L)_i
\right\}
+ 
\lambda'_{ijk}\left\{
-\overline{(d_R)_k}(u_L)_j (\widetilde{\ell}_L)_i
\right\}
+ {\rm h.c.}. 
\end{eqnarray}
The effective Lagrangian is given by 
\begin{eqnarray}
{\cal L^{\it s}_{\rm eff}} &=& -\frac{1}{4}
\sum_{k=1}^3
\frac{\lambda^*_{kji}\lambda'_{kab}}{m_{\widetilde{l}_{Lk}}^2}
\bar{\nu}_j (1+\gamma_5) \ell_i ~
\bar{d}_b (1-\gamma_5) u_a. 
\end{eqnarray}
From eq.~(\ref{decayconst}) and equations of motion for $u$- and $d$-quarks,  
we find 
\begin{eqnarray}
\langle 0| 
\bar{d}_b \gamma_5 u_a | P(q) \rangle 
= -i \frac{m_P^2}{m_{u_a}+m_{d_b}} f_P. 
\label{psdecay}
\end{eqnarray}
Using eq.~(\ref{psdecay}), we obtain the $s$-channel contribution as 
\begin{eqnarray}
(A_s^P)_{ij} &=& -\frac{1}{2\sqrt{2} m_{\ell_i}} 
 \frac{m_P^2}{m_{u_a}+m_{d_b}} 
\sum_{k=1}^3 
\frac{\lambda^*_{kji}\lambda'_{kab}}{m_{\widetilde{\ell}_{Lk}}^2}. 
\label{asp}
\end{eqnarray}
The charged Higgs contribution can be calculated from the 
interaction Lagrangian, 
\begin{eqnarray}
{\cal L} &=& V_{u_a d_b}^* 
\left\{
\frac{g m_{d_b}}{\sqrt{2}m_W} \tan\beta \overline{d_b} P_L u_a H^- 
+
\frac{g m_{u_a}}{\sqrt{2}m_W} \cot\beta \overline{d_b} P_R u_a H^- 
\right\}
\nonumber \\
&&~~~
+ 
\frac{g m_{\ell_i}}{\sqrt{2}m_W} \tan\beta \overline{\nu_i} P_R \ell_i
H^+ + {\rm h.c.}, 
\label{chhiggs}
\end{eqnarray}
where $g$ denotes the SU(2)$_L$ gauge coupling constant, and 
$\tan\beta \equiv \langle H_u \rangle/
\langle H_d \rangle$ is a ratio of the vacuum expectation values 
of two Higgs doublets $H_u$ (the weak hypercharge $Y=1/2$) 
and $H_d$ ($Y=-1/2$).  
We obtain the charged Higgs contribution $A_s^P$ from eq.~(\ref{chhiggs}) 
as 
\begin{eqnarray}
A^P_H &=& -G_F V_{u_a d_b}^* \frac{m_{d_b}}{m_{u_a} + m_{d_b}} 
\frac{m_P^2}{m_{H^-}^2}
\left(\tan^2\beta - \frac{m_{u_a}}{m_{d_b}}\right).  
\label{a_h}
\end{eqnarray}
Since leptons in the final state due to the charged Higgs 
exchange are flavor diagonal, the indices $i,j$ are suppressed in 
l.h.s. of eq.~(\ref{a_h}). 
Note that the overall sign of the r.h.s. of eq.~(\ref{a_h}) is opposite 
compared to the $W$-boson exchange (\ref{wboson}), the charged Higgs
boson contribution destructively interferes with the $W$-boson
contribution when $\tan^2\beta > m_{u_a}/m_{d_b}$.   
%

%
Next we summarize the experimental data and constraints on 
the parameter $r_P$ for $P=D_s$ and $B^+$. 
Comparison of the experimental results of leptonic decay of $D_s$  
meson is often presented in terms of the decay constant $f_{D_s}$. 
The world average of the experimental data of $D_s \to \tau \nu$ 
has been given by HFAG as~\cite{hfag:fds}
\begin{eqnarray}
\fds^{\rm exp} &=& 257.3 \pm 5.3~{\rm [MeV]}, 
\label{hfag2011}
\end{eqnarray}
while the HPQCD collaboration has updated their evaluation of the 
decay constant $\fds$ as~\cite{Davies:2010ip}
\begin{eqnarray}
\fds^{\rm SM} = 248 \pm 2.5~{\rm [MeV]}. 
\label{fdqcd2010}
\end{eqnarray}
The discrepancy between eqs.~(\ref{hfag2011}) and (\ref{fdqcd2010}) is 
$1.6\sigma$, and we find that the constraint on 
$r_{D_s}$-parameter as 
\begin{eqnarray}
r_{D_s} = 1.04 \pm 0.03. 
\label{rparamexp_ds}
\end{eqnarray}
%

%
For the leptonic decay $B^+ \to \tau \nu$ , 
the experimental data of the branching ratio have been given 
by Belle and BABAR~\cite{ikado,btau}. 
The average of the data and the SM prediction given by the UTfit 
collaboration~\cite{Bona:2009cj} is  
\begin{eqnarray}
{\rm Br}(B^+ \to \tau \nu)_{\rm exp} &=& 
(1.73 \pm 0.34)\times 10^{-4}, 
\label{expB}
\\
{\rm Br}(B^+ \to \tau \nu_\tau)_{\rm SM} 
&=& (0.84 \pm 0.11) \times 10^{-4}. 
\label{utfit}
\end{eqnarray}
The difference between (\ref{expB}) and (\ref{utfit}) is $2.5\sigma$ and
we find  
\begin{eqnarray}
r_{B^+} &=& 1.44 \pm 0.23. 
\label{rparamexp_bp}
\end{eqnarray}
%
%
%
%

%
\section{Numerical Study}\label{numerical}
As is shown in eq.~(\ref{rparam}), the new physics contributions to 
the leptonic decay of a pseudo scalar meson $P$ consist of the lepton 
flavor diagonal part $A_{ii}^P$ and off-diagonal part 
$A^P_{ij} (i\neq j)$. 
We study the contributions of $R$-parity violating interactions 
to the diagonal part $A_{ii}^P$~\cite{Aida:2010xi} in sec.~\ref{sss}, 
and the off-diagonal part $A^P_{ij} (i\neq j)$ taking account of 
the other decay processes in sec.~\ref{ss2}. 
In our analysis, we assume that the RPV couplings are real. 
We consider exchanges of 
the left-handed smuon and the right-handed sbottom 
in the $s$- and $t$-channel diagrams, respectively. 
Recall that the operator $L_i L_j E_k$ in eq.~(\ref{rparity})
is anti-symmetric for $i$ and $j$. 
This is why we consider the smuon instead of the stau in the 
$s$-channel diagram. 
Throughout in our numerical analysis, the sfermion masses are fixed at  
$100~{\rm GeV}$, and we adopt the central values of the 
following parameters~\cite{PDG2010} 
\begin{eqnarray}
|V_{cs}|&=&1.023\pm 0.036, 
~|V_{ub}|=(3.89\pm 0.44)\times 10^{-3}, 
~|V_{cb}|=(40.6\pm 1.3)\times 10^{-3}, 
\nonumber \\
m_{D_s} &=& 1968.47 \pm 0.33~{\rm MeV}, ~~
m_{B^+} = 5279.17 \pm 0.4~{\rm MeV}, ~~ 
m_{B^0} = 5279.5 \pm 0.5~{\rm MeV}. 
\nonumber \\
\end{eqnarray}

\subsection{
Constraints on the RPV couplings from the flavor diagonal final state 
}
\label{sss}
\begin{figure}[t]
 \begin{center}
  \includegraphics[width=6.5cm]{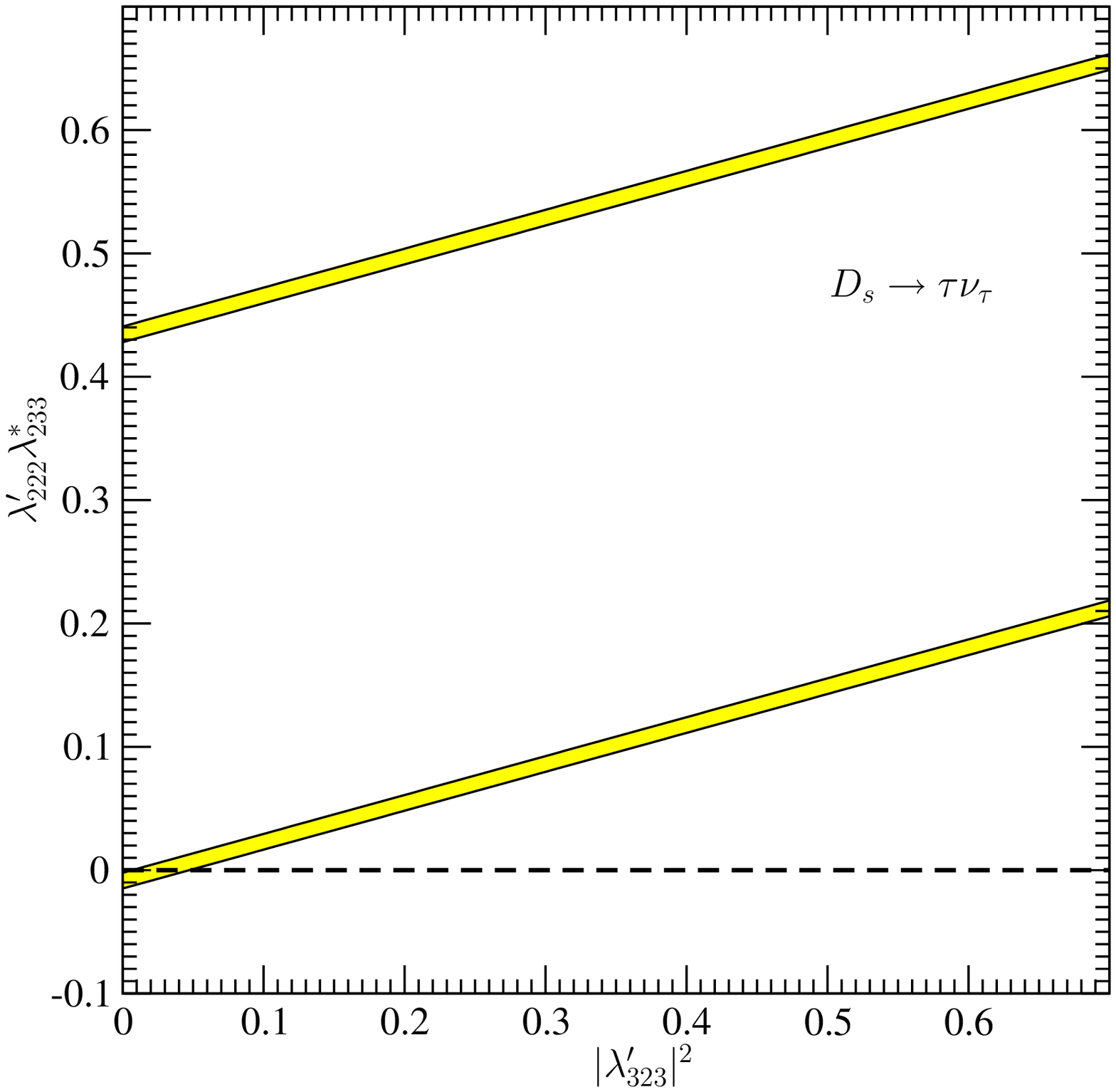}~~
  \includegraphics[width=7.2cm]{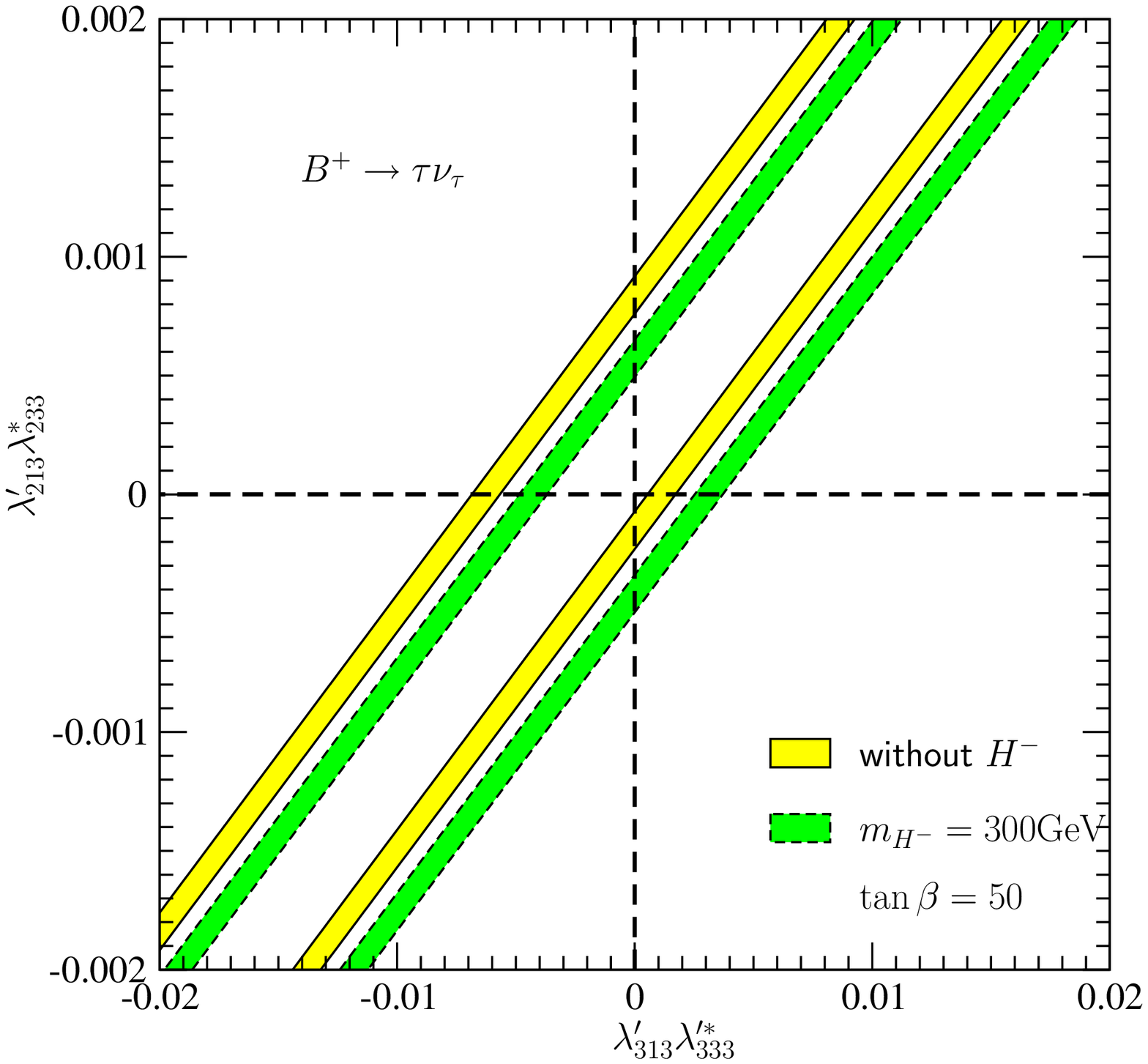}
\caption{
Constraints on the RPV couplings from 
$D_s \to \tau \nu$ (left)  
and $B^+ \to \tau \nu$ (right). The horizontal axis represents 
the RPV couplings for the $t$-channel while the vertical axis 
denotes the couplings for the $s$-channel diagram. 
The bands correspond to the 1-$\sigma$ allowed range for the 
$r_{D_s}$ (left) and $r_{B^+}$ (right) parameters, respectively. 
}
\label{fig_diag}
 \end{center}
\end{figure}
In this section, we focus on the final state where the lepton pair 
is flavor diagonal 
 ($\tau \nu_\tau$), i.e., we set $A_{ij}=0$ in eq.~(\ref{rparam}). 
The experimental data of both $D_s \to \tau \nu$ and 
$B^+ \to \tau \nu$ favor the positive interference between 
the SM $W$-boson exchange and the new physics contribution as shown 
in eqs.~(\ref{rparamexp_ds}) and (\ref{rparamexp_bp}). 
Due to the sign of r.h.s. in the $s$-channel slepton exchange 
(\ref{asp}) and the $t$-channel squark exchange (\ref{atp}), 
those contributions could be cancelled each other when the products of
RPV couplings in both $s$- and $t$-channel diagrams have the same sign. 
In Fig.~\ref{fig_diag}, we show the 1-$\sigma$ allowed region of the 
$r_{D_s}$-parameter (left) and the $r_{B^+}$-parameter (right) 
for $m_{\widetilde{\mu}_L}=m_{\widetilde{b}_R}=100{\rm GeV}$. 
The horizontal axis represents the RPV couplings for the $t$-channel 
diagram while the vertical axis denotes the couplings for the 
$s$-channel diagram. 
The charged Higgs boson contributions is negligibly small 
for $D_s \to \tau \nu$ but sizable for $B^+ \to \tau \nu$ 
when $\tan\beta$ is large. 
The dependence of the charged Higgs boson contributions to the 
$r_{B^+}$-parameter is shown in the right side of Fig.~\ref{fig_diag} 
for $\tan\beta=50$. 
The bands with solid line are obtained without the charged Higgs boson 
contribution while those with dashed line are obtained for 
$m_{H^-}=300~{\rm GeV}$, respectively. 
%

%
In each bands, the inner lines denote the 1-$\sigma$ lower bounds 
while the outers are the upper bounds. 
We find from the figures that the allowed region of $s$-channel
couplings shows positive correlations with the $t$-channel couplings. 
This is because the interference between the $s$- and $t$-channel 
contributions is destructive. 
For $D_s \to \tau \nu_\tau$, since the $t$-channel coupling is always 
positive ($|\lambda'_{323}|^2 \ge 0$), not only the magnitude but also 
the sign of $s$-channel coupling $\lambda'_{222}\lambda^*_{233}$ 
is strongly constrained. 
For $\lambda'_{222}\lambda^*_{233} \le 0$, 
the $t$-channel coupling $|\lambda'_{323}|^2$ should be smaller than 
$0.07$, and $-0.08 \simlt \lambda'_{222}\lambda^*_{233} \le 0$ 
is experimentally allowed in the 1-$\sigma$ level. 
For $B^+ \to \tau \nu_\tau$, 
the $s$- and $t$-channel couplings with opposite 
signs are strongly constrained. 
As can be seen in Fig.~\ref{fig_diag}, when the $t$-channel 
coupling is positive  
($\lambda'_{313}\lambda'^*_{333} \ge 0$), the negative $s$-channel
coupling is constrained to be  
$-0.0004 \simlt \lambda'_{213}\lambda^*_{233}\le 0$ when 
there is no charged Higgs contribution. 
Although the leptonic decays of $D_s$ and $B^+$ mesons are useful 
to constrain the sign of the relevant RPV couplings, 
the size of the couplings cannot be restricted because of the 
cancellation between the $s$- and $t$-channel diagrams. 
However, since several RPV couplings could be large simultaneously, 
it may lead to observation of productions or decay processes 
of SUSY particles due to the RPV interactions at LHC.

\subsection{Constraints on the RPV couplings from the final state with  
lepton flavor violation}
\label{ss2}
We have so far neglected the contributions of RPV interactions to 
the flavor off-diagonal part, $A^P_{ij}$ in eq.~(\ref{rparam}).  
In this subsection we examine $A^P_{ij}$ taking account of 
experimental bounds on the RPV couplings from 
the other lepton flavor violating processes. 
Since the lepton flavor off-diagonal terms in eq.~(\ref{rparam}) 
are always positive, one may expect that the deviation from the 
SM prediction could be explained by these terms when 
the flavor diagonal contribution $A^P_{ii}$ is negligible due to the  
cancellation between the $s$- and $t$-channel diagrams. 
We set, therefore, $A^P_{ii}=0$ in the following analysis. 
%

%
\subsubsection{$\dstaunu$}
\begin{figure}[h]
 \begin{center}
  \includegraphics[width=14cm]{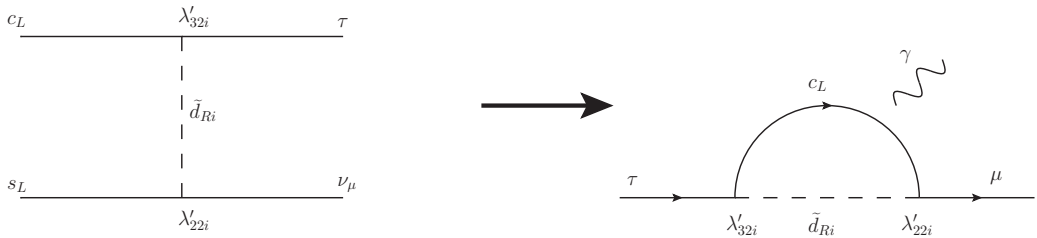}
\caption{
The Feynman diagram of the $t$-channel squark exchange in 
$\dstaunu$ (left) and the LFV process $\tau \to \mu \gamma$ (right). 
}
\label{fig_tgm}
 \end{center}
\end{figure}

First we examine $\dstaunu$. 
As shown in Fig.~\ref{fig_tgm}, 
replacing $(s_L, \nu_\mu)$ in the $t$-channel diagram by 
$(c_L, \mu_L)$, 
we obtain the Feynman diagram for the lepton flavor violating process 
$\tau \to \mu \gamma$. 
For $\tau \to \mu \gamma$, the relevant RPV couplings are 
$\lambda'_{32i}\lambda'_{22i}$ where $i$ denotes the
generation index of the right-handed down squark. 
Constraints on $\lambda'_{32i}\lambda'_{22i}$ from $\tau \to \mu\gamma$ 
have been studied in ref.~\cite{Bhattacharyya:2009hb} 
using the experimental data given by Belle~\cite{Hayasaka:2007vc} as, 
\begin{eqnarray}
{\rm Br}(\tau \to \mu\gamma) < 4.5\times 10^{-8}. 
\end{eqnarray}
They found the constraint on the RPV couplings 
$\lambda'_{32i}\lambda'_{22i}< 9 \times 10^{-2}$ for 
$m_{\widetilde{d}_R}=300~{\rm GeV}$. 
Since we adopt $m_{\widetilde{d}_R}=100~{\rm GeV}$ as a reference value 
in our numerical study, the constraint on the RPV coupling can be read
as 
\begin{eqnarray}
\lambda'_{32i}\lambda'_{22i}< 1 \times 10^{-2}. 
\label{rpvbound1101}
\end{eqnarray}
The bound (\ref{rpvbound1101}) will be as small as 
$\lambda'_{32i}\lambda'_{22i}< 4 \times 10^{-3}$ 
for ${\rm Br}(\tau \to \mu\gamma) < 1 \times 10^{-8}$ 
which is expected to be achieved at future super B-factory with 
the integrated luminosity $5{\rm ab}^{-1}$~\cite{Abe:2010sj}. 
\begin{figure}[t]
 \begin{center}
  \includegraphics[width=7cm]{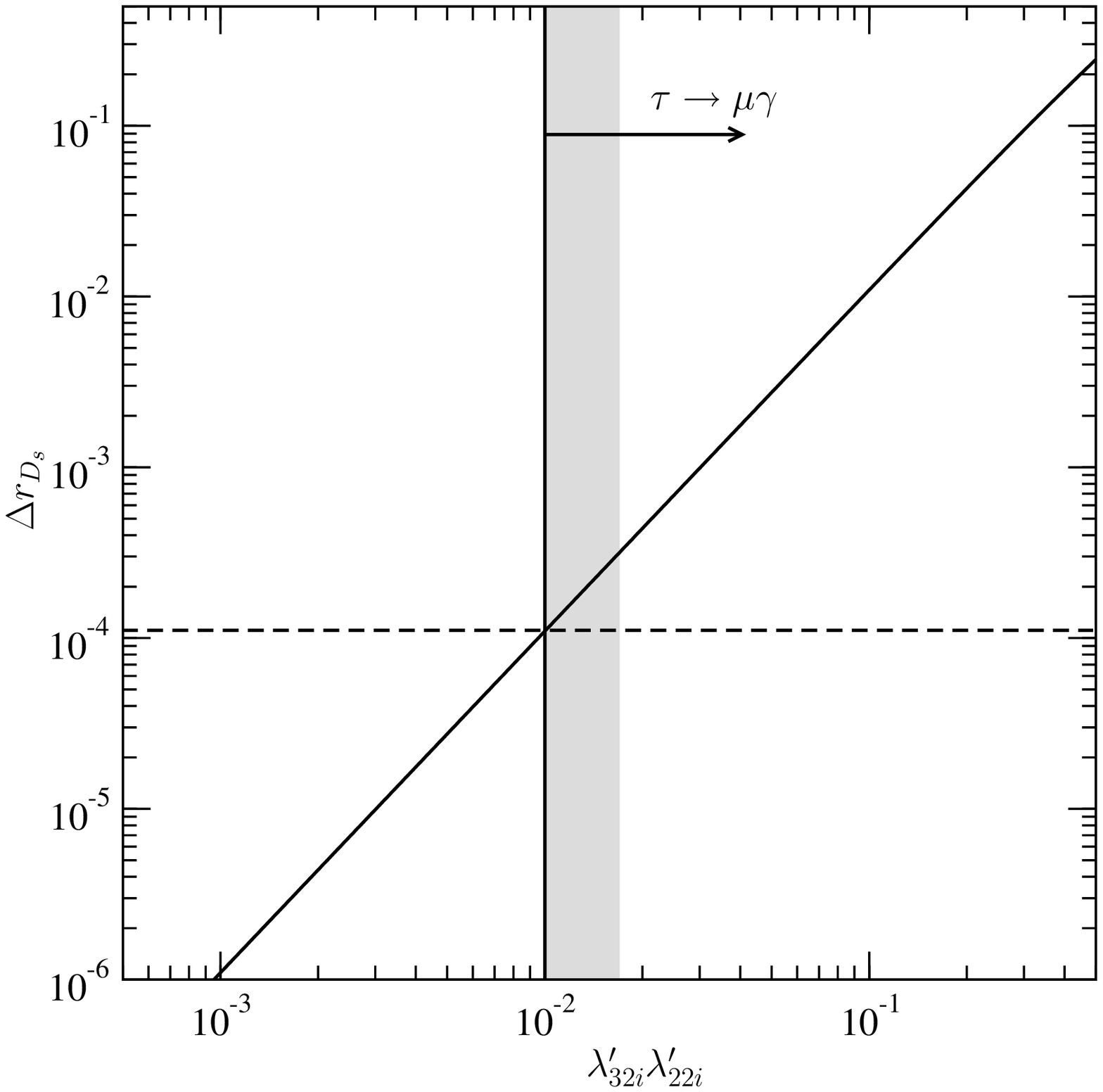}~~~~~
  \includegraphics[width=7cm]{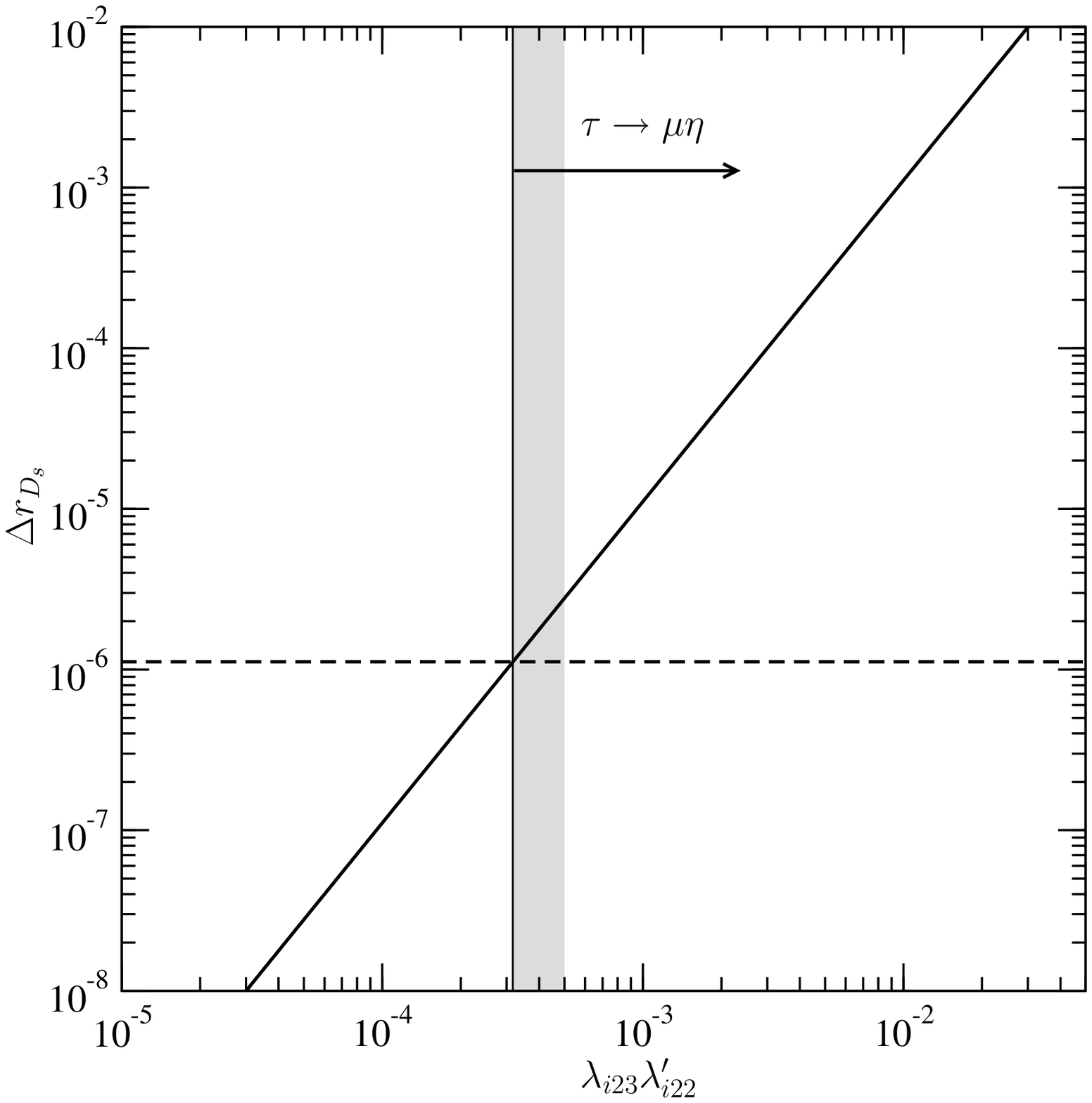}
\caption{
The flavor off-diagonal contributions to the $r_{D_s}$-parameter 
as function of RPV couplings. The vertical line denotes the 
experimental bound on the RPV couplings from $\tau \to \mu \gamma$.
\label{fig_lfv_ds}
}
 \end{center}
\end{figure}
%

%
We introduce a parameter $\Delta r_P (P=D_s, B^+)$ to estimate the 
contributions of the RPV interactions as 
\begin{eqnarray}
\Delta r_P \equiv r_P - \left(r_P \right)_{\rm SM},~~(P=D_s, B^+), 
\end{eqnarray}
where $(r_P)_{\rm SM}=1$. 
In Fig.~\ref{fig_lfv_ds} (left), the parameter $\Delta \rds$ is shown  
as a function of the RPV coupling $\lambda'_{32i}\lambda'_{22i}$. 
The vertical line denotes the experimental bound on 
$\lambda'_{32i}\lambda'_{22i}$ given in eq.~(\ref{rpvbound1101}). 
When $\lambda'_{32i}\lambda'_{22i} = 1\times 10^{-2}$, 
the magnitude of $\Delta \rds$-parameter is $\sim 10^{-4}$. 
Since the experimental bound on the $\rds$-parameter is 
$r_{D_s} = 1.04 \pm 0.03$ (\ref{rparamexp_ds}), 
$\Delta\rds \simlt 10^{-4}$ is small enough so that 
the contributions from the RPV coupling 
$\lambda'_{32i}\lambda'_{22i}$ is negligible. 

\begin{figure}[h]
 \begin{center}
  \includegraphics[width=14cm]{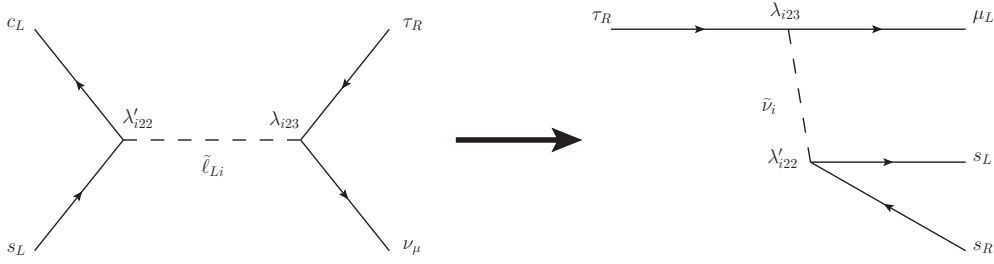}
\caption{
The Feynman diagram of the $s$-channel slepton exchange in 
$D_s \to \tau \nu_\mu$ (left) and the LFV process 
$\tau \to \mu \eta$ (right). 
}
\label{fig_tme}
 \end{center}
\end{figure}
A product of RPV couplings $\lambda_{i23}\lambda'_{i22}$ which appears 
in the $s$-channel diagram of $D_s \to \tau \nu_\mu$ also lead to the 
lepton flavor violating process $\tau \to \mu \eta$ 
as shown in Fig.~\ref{fig_tme}. 
The constraint on $\lambda_{i23}\lambda'_{i22}$ has been studied by 
Li et al.~\cite{Li:2005rr} as 
$\lambda_{i23}\lambda'_{i22}< 8.03\times 10^{-4}$ for
$m_{\widetilde{\ell}_{Li}}=100{\rm GeV}$ 
from the experimental limit 
${\rm Br}(\tau \to \mu\eta) < 1.5 \times 10^{-7}$ 
given by Belle in 2005~\cite{Enari:2005gc}.  
We update the bound given in ref.~\cite{Li:2005rr} using the new 
data in 2010 by Belle~\cite{hayasaka:2010}  
\begin{eqnarray}
{\rm Br}(\tau \to \mu\eta) < 2.3 \times 10^{-8}, 
\end{eqnarray}
and find that the bound on the RPV couplings is stronger than a previous 
one as 
\begin{eqnarray}
\lambda_{i23}\lambda'_{i22}< 3.1\times 10^{-4}. 
\label{i23i22a}
\end{eqnarray} 
We show $\Delta \rds$ as a function of the RPV couplings 
$\lambda_{i23}\lambda'_{i22}$ in Fig.~\ref{fig_lfv_ds} (right). 
The vertical line denotes the experimental limit on 
$\lambda_{i23}\lambda'_{i22}$ from ${\rm Br}(\tau \to \mu\eta)$ 
shown in eq.~(\ref{i23i22a}). 
It is easy to see that, taking account of the constraint on 
RPV coupling (\ref{i23i22a}), the magnitude of $\Delta \rds$-parameter  
is at most $\sim 10^{-5}$, which is negligibly small 
as compared to the experimental uncertainty (\ref{rparamexp_bp}). 

We conclude that, in $\dstaunu$,  the contributions of the RPV 
interactions to the lepton flavor off-diagonal part 
$A_{ij}$ in eq.~(\ref{rparam}) is highly suppressed from 
the constraints on the other LFV processes,  
$\tau \to \mu \gamma$ and $\tau \to \mu \eta$.  
The deviation from the SM contribution in the $\rds$-parameter is 
predicted to be smaller than $10^{-4}$ ($t$-channel) or $10^{-5}$
($s$-channel) so that we can safely neglect these contributions. 

\begin{figure}[t]
 \begin{center}
  \includegraphics[width=7cm]{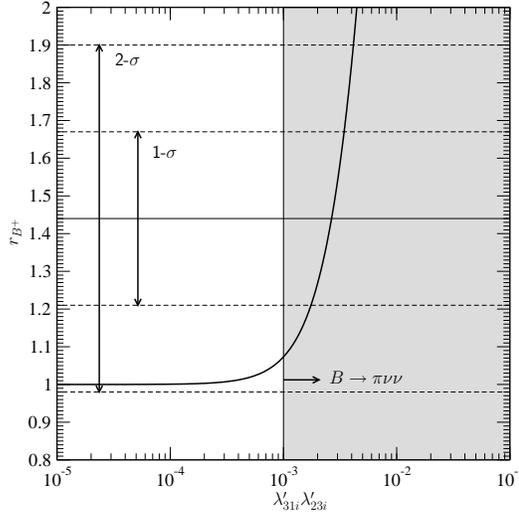}
  \caption{
The $r_{B^+}$-parameter as a function of the RPV couplings 
in the $t$-channel diagram, $\lambda'_{31i}\lambda'_{23i}$ 
for the slepton mass $m_{\widetilde{\ell}_{iL}}=100~{\rm GeV}$. 
The contributions from the flavor diagonal part in the final 
leptons are set to be zero, i.e., $A_{ii}^{B^+}=0$ in eq.~(\ref{rparam}). 
\label{fig_bdeay_tch1}
}
 \end{center}
\end{figure}
\subsubsection{$\bptaunu$}
For $\bptaunu$, the relevant RPV couplings in $A_{3j}^{B^+}$ 
(\ref{rparam}) are 
$\lambda'_{31i}\lambda'_{j3i}$ and $\lambda_{ij3}\lambda'_{i13}$ 
for the $t$- and the $s$-channel diagrams, respectively. 
Note that the indices $i$ and $j$ denote the flavor of 
slepton $\widetilde{\ell}_{Li}$ and neutrino $\nu_j$. 
Hereafter we consider $j=2$ in our analysis for simplicity. 

The couplings in the $t$-channel diagram,
$\lambda'_{31i}\lambda'_{23i}$,  
also appear in the semileptonic decays of $b$-quark such as 
$b \to u \ell \nu$ or $b \to d \nu \nu$. Since the latter is 
a typical process due to the flavor changing neutral current, 
the SM contribution is given by the 1-loop diagram while the RPV 
interactions lead to the same process at the tree level. 
Therefore strong constraints can be expected 
from $b \to d \nu \nu$. 

Constraints on the RPV coupling $\lambda'_{31i}\lambda'_{23i}$ from 
$B^+ \to \pi^+ \nu \bar{\nu}$ have been studied in ref.~\cite{Kim:2009mp}.   
The authors in ref.~\cite{Kim:2009mp} found the constraint on the 
RPV coupling $\lambda'_{31i}\lambda'_{23i} < 2.5 \times 10^{-2}$ for  
$m_{\widetilde{d}_R}=500{\rm GeV}$. 
It can be read, for $m_{\widetilde{d}_R}=100{\rm GeV}$, as 
\begin{eqnarray}
\lambda'_{31i}\lambda'_{23i} < 1.0 \times 10^{-3}. 
\end{eqnarray}
We depict the $\rbp$-parameter as a function of the RPV couplings 
$\lambda'_{31i}\lambda'_{23i}$ for the squark mass 
$m_{\widetilde{d}_R}=100{\rm GeV}$ in Fig.~\ref{fig_bdeay_tch1}. 
In the figure, we set the contributions to the $\rbp$-parameter due 
to the RPV interactions in the diagonal part of the lepton flavor 
to be zero, i.e., $A_{ii}^{B^+}=0$ in eq.~(\ref{rparam}). 
It is easy to see that, taking account of the bound from 
$B^+ \to \pi^+ \nu \bar{\nu}$, 
the contributions from the couplings $\lambda'_{31i}\lambda'_{23i}$ 
are not enough to explain the data of $\rbp$-parameter in the 1-$\sigma$
level, $r_{B^+} = 1.44 \pm 0.23$ in eq.~(\ref{rparamexp_bp}), 
though the contributions slightly increase the $\rbp$-parameter
from unity.  

Next we study contributions of the RPV interactions in the $s$-channel 
diagram. 
The combination of the RPV couplings $\lambda_{ij3}\lambda'_{i13}$ 
in the $s$-channel diagram of $B^+ \to \tau \nu_j~(j=1,2~{\rm
or}~e,\mu)$ can also  
contribute to the lepton flavor violating processes 
$B^0 \to \tau^\pm e^\mp$ or $B^0 \to \tau^\pm  \mu^\mp$. 
The decay rate of $B^0 \to \ell^\pm_i  \ell^\mp_j$ is given by 
\begin{eqnarray}
\Gamma(B^0 \to \ell^\pm_i  \ell^\mp_j)
&=& \frac{1}{128\pi}
\frac{|\lambda_{kij}\lambda'_{k13}|^2}{m_{\widetilde{\nu}_k}^4}
f_B^2 \frac{m_B^5}{m_b^2}
\left(1-\frac{m_{\ell_i}^2}{m_B^2}\right)^2
\nonumber \\
&\approx& 2.2 \times 10^{-10} 
\frac{|\lambda_{kij}\lambda'_{k13}|^2}
{\left(\frac{m_{\tilde{\nu}_k}}{100{\rm GeV}}\right)^4}, 
\end{eqnarray}
where $m_{\ell_i} \gg m_{\ell_j}$ is assumed. 
The branching ratio of $B^0 \to \ell^\pm_i  \ell^\mp_j$ is obtained 
by 
\begin{eqnarray}
{\rm Br}(B^0 \to \ell^\pm_i \ell^\mp_j) 
= \frac{\Gamma(B^0 \to \ell^\pm_i \ell^\mp_j) }{\Gamma(b \to c e \bar{\nu})}
{\rm Br}(b \to c e \bar{\nu}). 
\end{eqnarray}
The experimental bounds on 
${\rm Br}(B^0 \to \tau^\pm e^\mp)$ and 
${\rm Br}(B^0 \to \tau^\pm \mu^\mp)$  
are summarized by HFAG as~\cite{hfag:2010summer}
\begin{subequations}
\begin{eqnarray}
{\rm Br}(B^0 \to \tau^\pm e^\mp) &<& 28 \times 10^{-6}, 
\label{taue}
\\
{\rm Br}(B^0 \to \tau^\pm \mu^\mp) &<& 22 \times 10^{-6}. 
\label{taumu}
\end{eqnarray} 
\label{hfagdata}
\end{subequations}
\begin{figure}[t]
 \begin{center}
  \includegraphics[width=7cm]{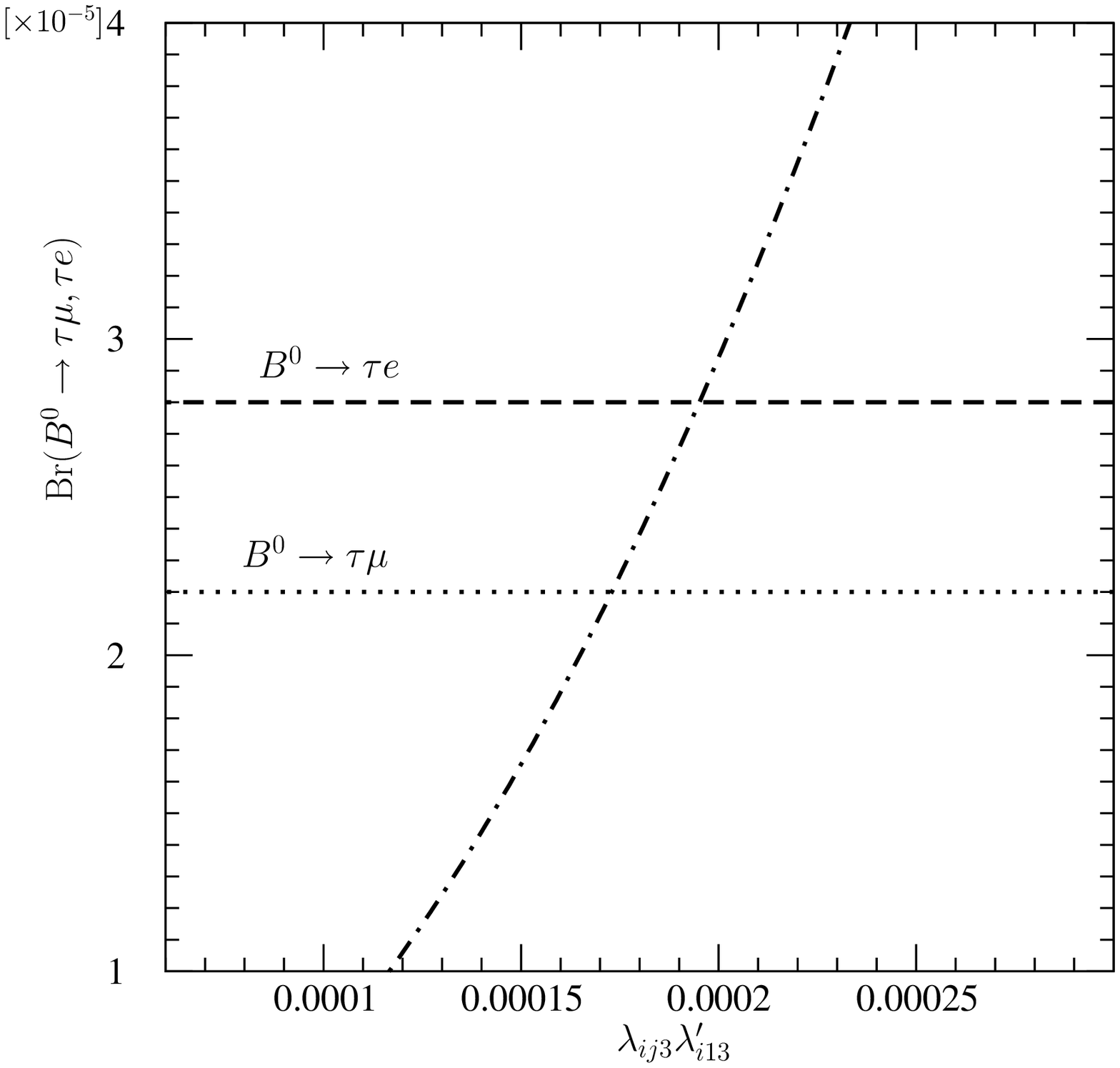}~~~~~
  \includegraphics[width=7cm]{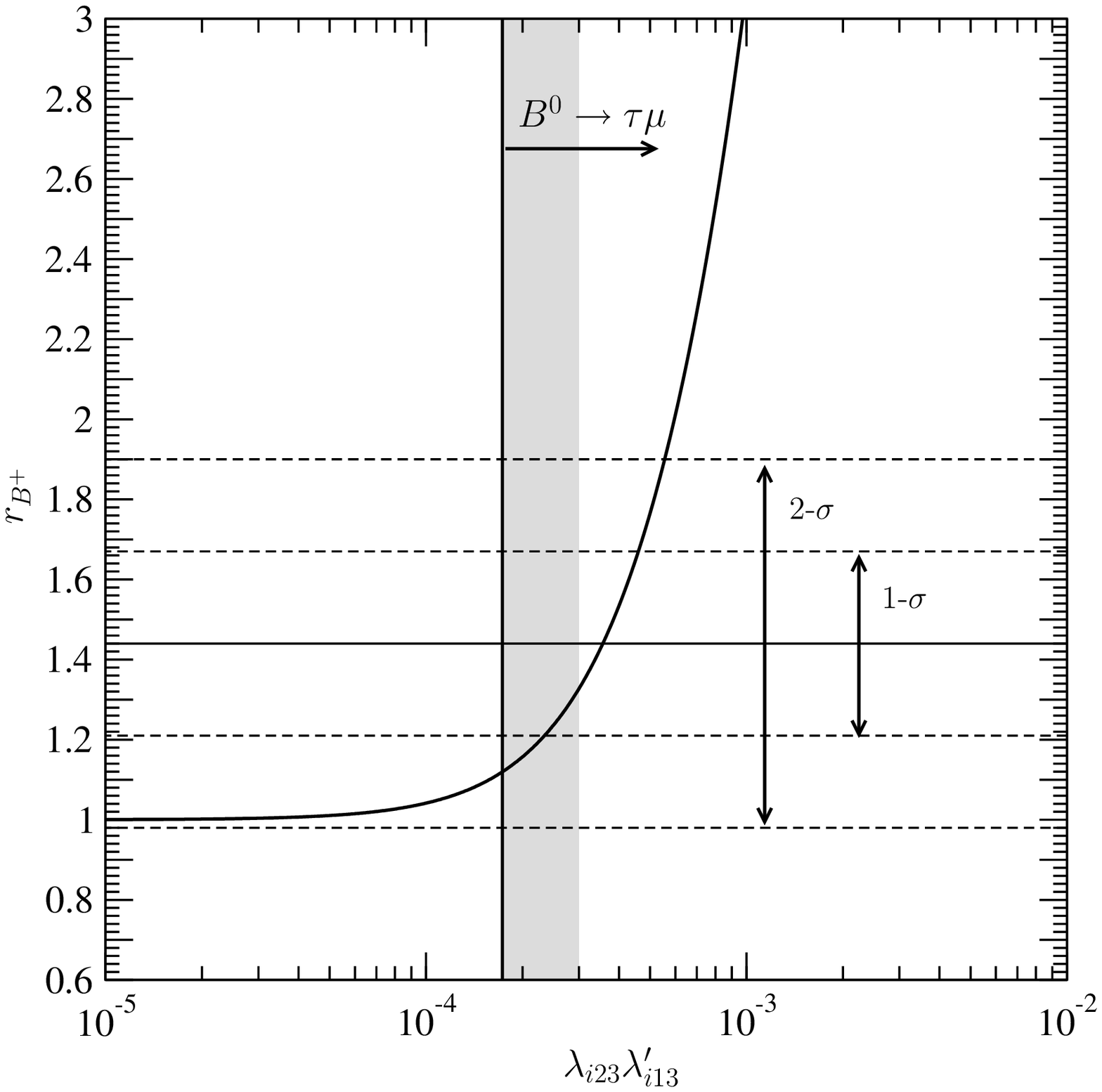}
  \caption{
(Left) 
The branching ratio ${\rm Br}(B^0 \to \ell_i^\pm \ell_j^\mp)$ as a 
function of the RPV couplings for the sneutrino mass 
$m_{\widetilde{\nu}}=100{\rm GeV}$. 
The horizontal lines denote the experimental bounds on 
$B^0 \to \tau^\pm e^\mp$ (dashed) and $B^0 \to \tau^\pm \mu^\mp$ 
(dotted), respectively.  
(Right) 
The $\rbp$-parameter as a function of the RPV couplings. 
The vertical line represents the upper limit on the 
RPV couplings from the experimental data of  
${\rm Br}(B^0 \to \tau^\pm \mu^\mp)$. 
\label{fig_bdeay_br}
}
\end{center}
\end{figure}
We show the branching ratio of $B^0 \to \ell_i^\pm \ell_j^\mp$ 
as a function of the RPV couplings for the $s$-channel diagram 
of $\bptaunu$ in Fig.~\ref{fig_bdeay_br}~(left). 
The sneutrino mass is fixed at $100{\rm GeV}$. 
In the figure, two horizontal lines denote the experimental 
upper bound of $B^0 \to \tau^\pm e^\mp$ (dashed) and 
$B^0 \to \tau^\pm \mu^\mp$ (dotted), respectively.  
Due to the experimental bounds on $B^0 \to \tau^\pm e^\mp$ and 
$B^0 \to \tau^\pm \mu^\mp$ in eqs.~(\ref{taue}) and (\ref{taumu}), 
the upper limits of the RPV couplings slightly differ 
between $\lambda_{i23}\lambda'_{i13}$ and 
$\lambda_{i13}\lambda'_{i13}$. 
From the figure, we find constraints on the RPV couplings as 
\begin{subequations}
\begin{eqnarray}
\lambda_{i23}\lambda'_{i13} &\simlt& 1.7 \times 10^{-4}, 
\label{lambdataumu}
\\
\lambda_{i13}\lambda'_{i13} &\simlt& 1.9 \times 10^{-4}. 
\label{lambdamue}
\end{eqnarray}
\end{subequations}

Now we are ready to study contribution to the $\rbp$-parameter 
through the RPV interactions with couplings 
$\lambda_{i23}\lambda'_{i13}$ and $\lambda_{i13}\lambda'_{i13}$ 
taking account of the experimental constraints from 
$B^0 \to \ell_i^\pm \ell^\mp_j$, (\ref{lambdataumu}) and 
(\ref{lambdamue}).  
We show the $\rbp$-parameter as a function of the RPV couplings 
$\lambda_{i23}\lambda'_{i13}$ 
in Fig.~\ref{fig_bdeay_br}~(right). 
The 1- and 2-$\sigma$ bounds of the $\rbp$-parameter are shown 
by the horizontal dashed lines as indicated in the figure. 
The vertical line denotes the constraints on the RPV couplings 
from the branching ratio of $B^0 \to \tau^\pm \mu^\mp$. 
We find that, taking account of the experimental data of 
${\rm Br}(B^0 \to \tau^\pm \mu^\mp)$, the constraints on $\rbp$-parameter 
given by the lepton flavor off-diagonal coupling 
$\lambda_{i23}\lambda'_{i13}$ is $\rbp \simlt 1.15$, 
which is smaller than the 1-$\sigma$ lower bound, $\rbp = 1.21$. 
However, $\rbp$ may be enhanced when we add contributions from 
$\lambda_{i23}\lambda'_{i13}$ and $\lambda_{i13}\lambda'_{i13}$. 
Under the constraints given in eqs.~(\ref{lambdataumu}) and (\ref{lambdamue}), 
we find the $\rbp$-parameter could be at max 1.22 which is slightly
larger than the 1-$\sigma$ lower bound, $\rbp=1.21$. 
The deviation in the $\rbp$-parameter from the SM expectation can be, 
therefore, explained by the RPV couplings 
$\lambda_{i23}\lambda'_{i13}$ and $\lambda_{i13}\lambda'_{i13}$ 
which lead to the lepton flavor off-diagonal final state 
in $\bptaunu$ without conflicting the other LFV processes such as 
$B^0 \to \tau^\pm \mu^\mp, \tau^\pm e^\mp$.

\section{Summary}
\label{summary}
We have studied the contributions from $R$-parity violating interactions
to $D_s \to \tau \nu$ and $B^+ \to \tau \nu$. 
Owing to the latest Lattice QCD calculation of the decay constant 
$\fds$, the difference between the data and the SM expectation is 
only 1.6$\sigma$ in $\dstaunu$ while there is still 2.5$\sigma$
deviation in $\bptaunu$. 

The $R$-parity violating interactions could contribute to the leptonic 
decays through the slepton exchange in the $s$-channel diagram and 
the down-squark exchange in the $t$-channel diagram. 
In addition, the RPV interactions predict the final states 
not only the flavor diagonal lepton pair ($\tau \nu_\tau$) but also 
flavor off-diagonal pair ($\tau \nu_i,~i=e,\mu$). 
In the flavor diagonal ($\tau\nu_\tau$) case, 
the supersymmetric contributions could  interfere with the SM $W$-boson
contribution either constructively or  
destructively, see eq.~(\ref{rparam}). 
We showed that, 
since the supersymmetric contributions consist of the $s$-channel 
and $t$-channel diagrams, the interference between them 
is also either constructive or destructive, owing to a choice of 
relative sign of the RPV couplings between two diagrams. 
It was also shown that 
the experimentally allowed region of the RPV couplings 
are found to be strongly restricted when the relative sign of RPV
couplings between two diagrams is opposite. 
On the other hand, 
when the RPV couplings in $s$- and $t$-channel diagrams have the same
sign, these couplings are allowed to be large simultaneously. 

In the flavor off-diagonal case, as shown in eq.~(\ref{rparam}), 
the new physics contributions $A_{ij}$ to the leptonic decay 
always interfere with the SM contribution constructively. 
The RPV interactions in $A_{ij}^{D_s}$ also contribute to 
the LFV processes $\tau \to \mu \gamma$ and $\tau \to \mu \eta$.  
Taking account of experimental bounds on the RPV couplings from these 
decay processes, we found that $A_{ij}^{D_s}$ is highly suppressed 
and the contribution is smaller than the experimental uncertainty by two 
or more orders of magnitude. 
The RPV couplings in the $t$-channel diagram of $B^+ \to \tau \nu_\mu$ 
are constrained from $B^+ \to \pi^+ \nu \bar{\nu}$ so that the
$t$-channel contribution to $A_{ij}^{B^+}$ cannot explain the
2.5$\sigma$ discrepancy between the data and the SM prediction 
in the leptonic decay of the $B^+$ meson. 
The RPV couplings in the $s$-channel diagram also contribute to 
the LFV processes $B^0 \to \tau^\pm \mu^\mp, \tau^\pm e^\mp$. 
We found that, taking account of the experimental bound on the RPV
couplings $\lambda_{i23}\lambda'_{i13}$ 
from $B^0 \to \tau^\pm \mu^\mp$, the $r_{B^+}$-parameter is 
given as $r_{B^+} \simlt 1.15$ which is smaller than the 
1-$\sigma$ lower bound ($r_{B^+} = 1.44 \pm 0.23$). 
However, taking a sum of contributions from 
$\lambda_{i23}\lambda'_{i13}$ in $B^+ \to \tau \nu_\mu$ 
and 
$\lambda_{i13}\lambda'_{i13}$ in 
$B^+ \to \tau \nu_e$, the $r_{B^+}$-parameter could be at max 
1.22 which is consistent with the 1-$\sigma$ lower bound.

\end{document}